\documentstyle[aps,epsf,rotate,subfigure]{revtex}

\begin{document}

\draft
\title{Density-dependent diffusion in the periodic Lorentz gas}
\author{Rainer Klages$^{1}$ and Christoph Dellago$^{2}$}
\address{$^1$Center for Nonlinear Phenomena and Complex Systems,
Universit\'{e} Libre de Bruxelles,\\ 
Campus Plaine CP 231, Blvd du Triomphe, B-1050 Brussels, Belgium;
e-mail: rklages@ulb.ac.be\\ 
$^2$Department of Chemistry, University of Rochester, Rochester, NY
14627; \\ 
e-mail: dellago@chem.rochester.edu}
\date{\today}
\maketitle
\begin{abstract}
We study the deterministic diffusion coefficient of the
two-dimensional periodic Lorentz gas as a function of the density of
scatterers. Results obtained from computer simulations are compared to
the analytical approximation of Machta and Zwanzig
[Phys.Rev.Lett.~{\bf 50},~1959~(1983)] showing that their argument is
only correct in the limit of high densities. We discuss how the
Machta-Zwanzig argument, which is based on treating diffusion as a
Markovian hopping process on a lattice, can be corrected
systematically by including microscopic correlations. We furthermore
show that, on a fine scale, the diffusion coefficient is a non-trivial
function of the density. We finally argue that, on a coarse scale and
for lower densities, the diffusion coefficient exhibits a
Boltzmann-like behavior, whereas for very high densities it crosses
over to a regime which can be understood qualitatively by the
Machta-Zwanzig approximation.\\[2ex]
\end{abstract}

\section{Introduction}
One of the central themes in the theory of chaotic transport is the
problem of deterministic diffusion. Over the past several years,
deterministic diffusion coefficients have been computed for a variety
of low-dimensional model systems. The most elementary ones are chains
of one-dimensional maps \cite{GF2,GeNi82,SFK,BaKl97}, which have been studied 
with cycle expansion methods\cite{Art1,Tse94,Chen95,Cvit95}. By applying 
techniques based on Markov partitions it has been found
that the deterministic diffusion coefficient of a simple piecewise
linear one-dimensional map is a fractal function of the control
parameter \cite{RKD,RKdiss}. Long-range dynamical correlations in the
microscopic chaotic scattering processes of the system are 
responsible for this surprising behavior
\cite{RKdiss,dcrc,KlDo99}. More complicated models exhibiting
deterministic diffusion are two-dimensional maps such as multi-Bakers
\cite{PG1,TG2,VTB97,GaKl}, cat maps \cite{DMP89},
standard maps \cite{ReWi80}, and related models
\cite{Lebo98}. Also in these models the diffusion
coefficients are fractal \cite{GaKl}, or are highly
nontrivial functions of a control parameter
\cite{DMP89,ReWi80,Lebo98}.

These simple maps share fundamental physical and mathematical
properties of more complicated dynamical systems such as the periodic
Lorentz gas \cite{Gasp}. For this model, the diffusion coefficient
 has been computed with different
methods: in Refs.\ \cite{MaZw83,BarEC,MaMa97} it has been obtained
via the Green-Kubo formula on the basis of computer
simulations, in Refs.\ \cite{CvGS92,MoRo94} techniques of cycle and
periodic orbit expansions have been applied, and in Refs.\
\cite{GB1,GaBa95} the diffusion coefficient has been obtained from
escape rate methods as well as from the Hausdorff dimension of the
fractal repeller of the Lorentz gas.

In this paper we focus on the behavior of the diffusion coefficient in
the two-dimensional periodic Lorentz gas under variation of the
density of scatterers. In Section II we introduce the model, briefly
sketch the simple analytical approximation obtained by Machta and
Zwanzig \cite{MaZw83} and compare it to our numerical results obtained
from computer simulations. In Section III we explain how the
Machta-Zwanzig argument can be corrected systematically by taking
microscopic correlations into account. In Section IV we compare our
numerical diffusion coefficient to a simple Boltzmann approximation
and argue that, on a coarse scale, there is a dynamical crossover for
the diffusion coefficient as a function of the density.  More detailed
computer simulations show that the diffusion coefficient exhibits a
non-trivial fine structure as a function of the density, as is
discussed in Section V.  In Section VI we summarize our results and
relate them to analogous findings in one- and two-dimensional maps.

\section{Diffusion as a simple Markov process: Machta-Zwanzig
approximation and numerical results}

\begin{figure}
\epsfxsize=7cm
\centerline{\epsfbox{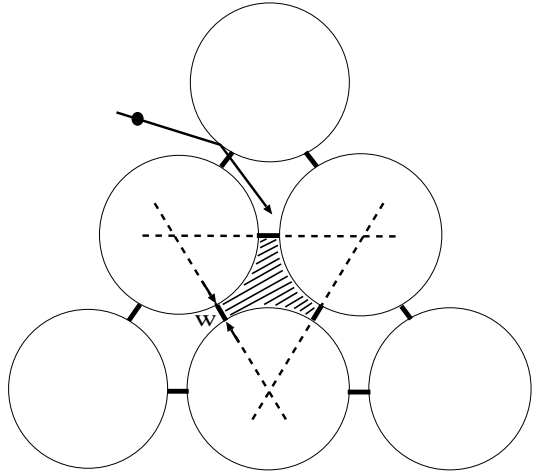}}
\vspace*{0.5cm}
\caption{Geometry of the periodic Lorentz gas. A point particle
scatters elastically at hard disks arranged on a triangular
lattice. The cross-hatching indicates a single triangular trapping
region. The smallest interdisk distance is denoted by $w$.}
\label{fig1}
\end{figure}

The periodic Lorentz gas is a standard model in the field of chaos and
transport \cite{Gasp,MaHo92,TGN98,DoVL}. It mimics classical diffusion
in a crystal and is isomorphic to a periodic system of two hard disks
per unit cell \cite{BuSp}. The geometry of the model is shown in Fig.\
\ref{fig1}: A point particle of mass $m$ moves with constant velocity
$v$ in an array of circular hard scatterers of radius $R$ arranged on
a triangular lattice. Upon collisions with the scatterers the particle
is reflected elastically. In the following we use units for which
$v=1$, $m=1$, and $R=1$. The lattice spacing of the disks is then
$2+w$, where $w$ is the smallest inter disk distance. The gap size $w$
is geometrically related to the number density $n$ of the disks by
\begin{equation}
n=2/[\sqrt{3}(2+w)^2] \quad . \label{eq:nd}
\end{equation}
The gap size $w$, or the number density of the disks $n$,
respectively, is the only control parameter and completely determines
the dynamical properties of the system. At close packing, i.e., $w=0$,
the moving particle is trapped in a single triangular region formed
between three disks (see Fig.\ \ref{fig1}). For
$0<w<w_{\infty}=4\sqrt{3}-2=0.3094$, the particle can move across the
entire lattice, but it cannot move collision-free for an infinite
time. For $w>w_{\infty}$ the particle can move arbitrarily far between
two collisions. Here, $w_{\infty}$ denotes the gap size at which the
particle first sees such an ``infinite horizon''. Bunimovich and Sinai
proved that for $0<w<w_{\infty}$ the system is ergodic, and that the
diffusion coefficient exists, while it diverges for $w>w_{\infty}$
\cite{BuSi1,BuSi2}.

\begin{figure}
\epsfxsize=7cm
\centerline{\rotate[r]{\epsfbox{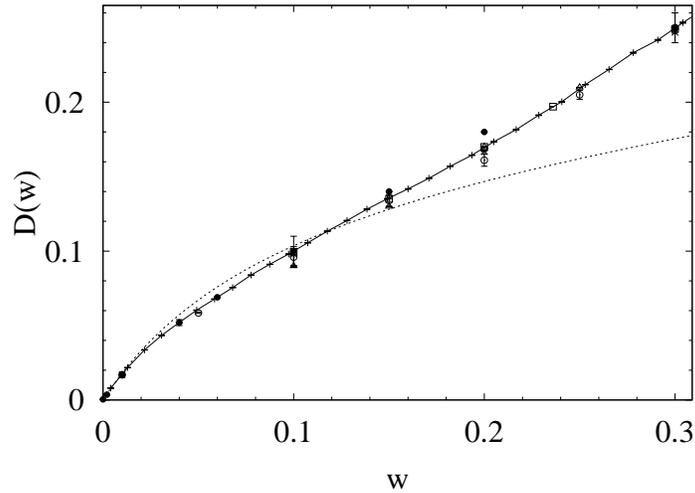}}}
\vspace*{0.3cm}
\caption{Diffusion coefficient $D$ as a function of the gap size
$w$. The dashed line represents the Machta-Zwanzig random walk model
Eq.\ (\ref{eq:mz}), the different symbols refer to single data points
obtained in the literature from various methods (see text). The
crosses connected with lines are our new results from
computer simulations. All of our data points have error bars smaller 
than the symbols. Error bars of the other data points have been
included as far as available.}
\label{fig2}
\end{figure}

In Ref.\ \cite{MaZw83} Machta and Zwanzig have derived a simple
analytical approximation for the diffusion coefficient
$D$. The basic idea is that diffusion can be treated as a Markovian
hopping process between the triangular trapping regions indicated in
Fig.\ \ref{fig1}. For this purpose, they have calculated the average
rate $\tau^{-1}$ at which a particle leaves such a trap.  According
to a simple phase space argument, this rate is determined by the
fraction of phase space available for leaving the trap divided by the
total phase space volume of the trap. Furthermore, for random walks on
two-dimensional isotropic lattices the diffusion coefficient is
$D=l^2/(4\tau)$, where $l=(2+w)/\sqrt{3}$ is the distance
between the centers of the traps. This leads to the Machta-Zwanzig
random walk approximation of the diffusion coefficient
\begin{equation}
D_{\rm MZ}=\frac{w(2+w)^2}{\pi[\sqrt{3}(2+w)^2-2\pi]} \quad . \label{eq:mz}
\end{equation}
This equation is shown as the dotted
line in Fig.\ \ref{fig2}. Included in this figure are single data points for the diffusion
coefficient obtained by various authors using different methods: The
filled circles are numerical results of Ref.\
\cite{MaZw83}, the stars are from Ref.\ \cite{BarEC}. These data
points have been obtained by employing the Green-Kubo formula, where
$D$ is determined from an integral over the velocity autocorrelation
function. The squares have been computed in Ref.\
\cite{MoRo94} by periodic orbit expansions, the empty circles
and triangles are from Ref.\ \cite{GaBa95}, where they have been
computed by applying the escape rate formalism, and via the fractal
dimension of the repeller of the Lorentz gas, respectively. The
crosses connected with lines represent our new data which we
have obtained from computer simulations by means of the Einstein
formula
\begin{equation}
D=\lim_{t\to\infty}\frac{\langle\Delta r^2(t)\rangle}{4t} \quad
, \label{eq:einst} 
\end{equation}
where the brackets indicate an ensemble average after time $t$.  We
calculated $\langle\Delta r^2(t)\rangle$ by averaging over long
trajectories. Depending on the density $n$ each trajectory contained
from $6\times 10^8$ to $3\times 10^9$ collisions. After a short
transient $\langle\Delta r^2(t)\rangle$ grows linearly with a slope of
$4D$. We determined the diffusion coefficient $D$ by fitting a
straight line to $\langle\Delta r^2(t)\rangle$ in the linear
regime. Though in principle equivalent with the Green-Kubo formalism,
the Einstein approach is numerically more efficient in the Lorentz
gas, where the equations of motion can be integrated exactly.

Fig.\ \ref{fig2} shows that the Machta-Zwanzig approximation Eq.\
(\ref{eq:mz}) is only correct in the limit of small gap sizes $w\to
0$. For larger values of $w$ and up to approximately $w<0.1$ Eq.\
(\ref{eq:mz}) clearly overestimates the exact diffusion coefficient,
whereas for $w > 0.1$ it significantly underestimates diffusion. In
the following section, we will discuss the physical reason for this
failure of the Machta-Zwanzig approximation and how it can be
corrected. Note also that for $w\to w_{\infty}$ there is no evidence
for any singularity in the diffusion coefficient reminiscent of
critical behavior. In Ref.\ \cite{MaMa97}, the same observation has
been made based on the analysis of velocity autocorrelation
functions. For explaining this lack of criticality the authors have
argued that this transition rather represents a simple geometric
feature of the system than being a consequence of long-range
correlations.

\section{Correction of the Machta-Zwanzig approximation: correlated
microscopic scattering}
\subsection{Collisionless flights across a trap}
\begin{figure}
\begin{center}
\epsfxsize=5.7cm
\subfigure[\hspace*{4.8cm}]{\rotate[r]{\epsfbox{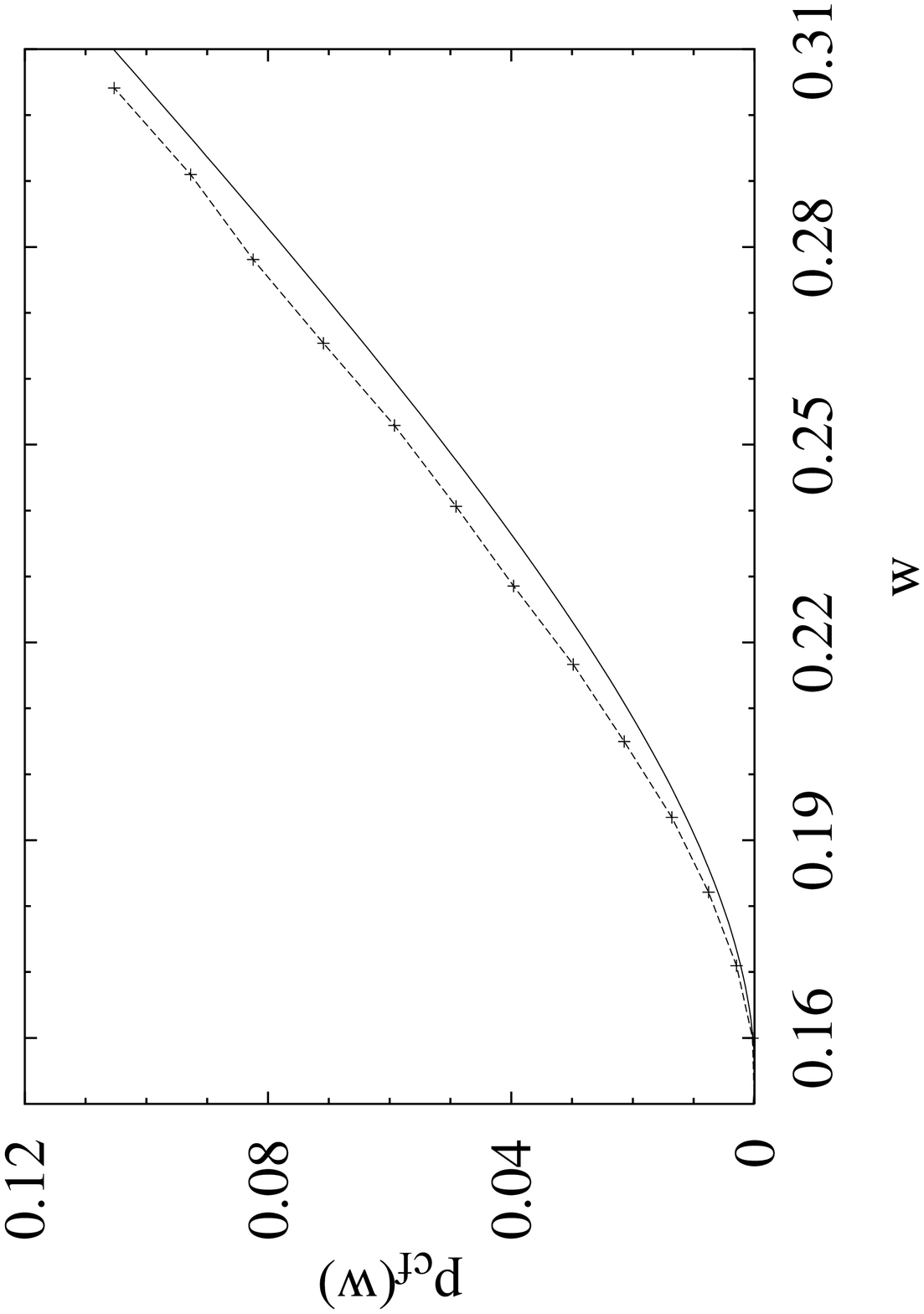}\hspace*{-0.5cm}}}
\hspace*{0.3cm}
\epsfxsize=5.5cm
\subfigure[\hspace*{5.5cm}]{\rotate[r]{\epsfbox{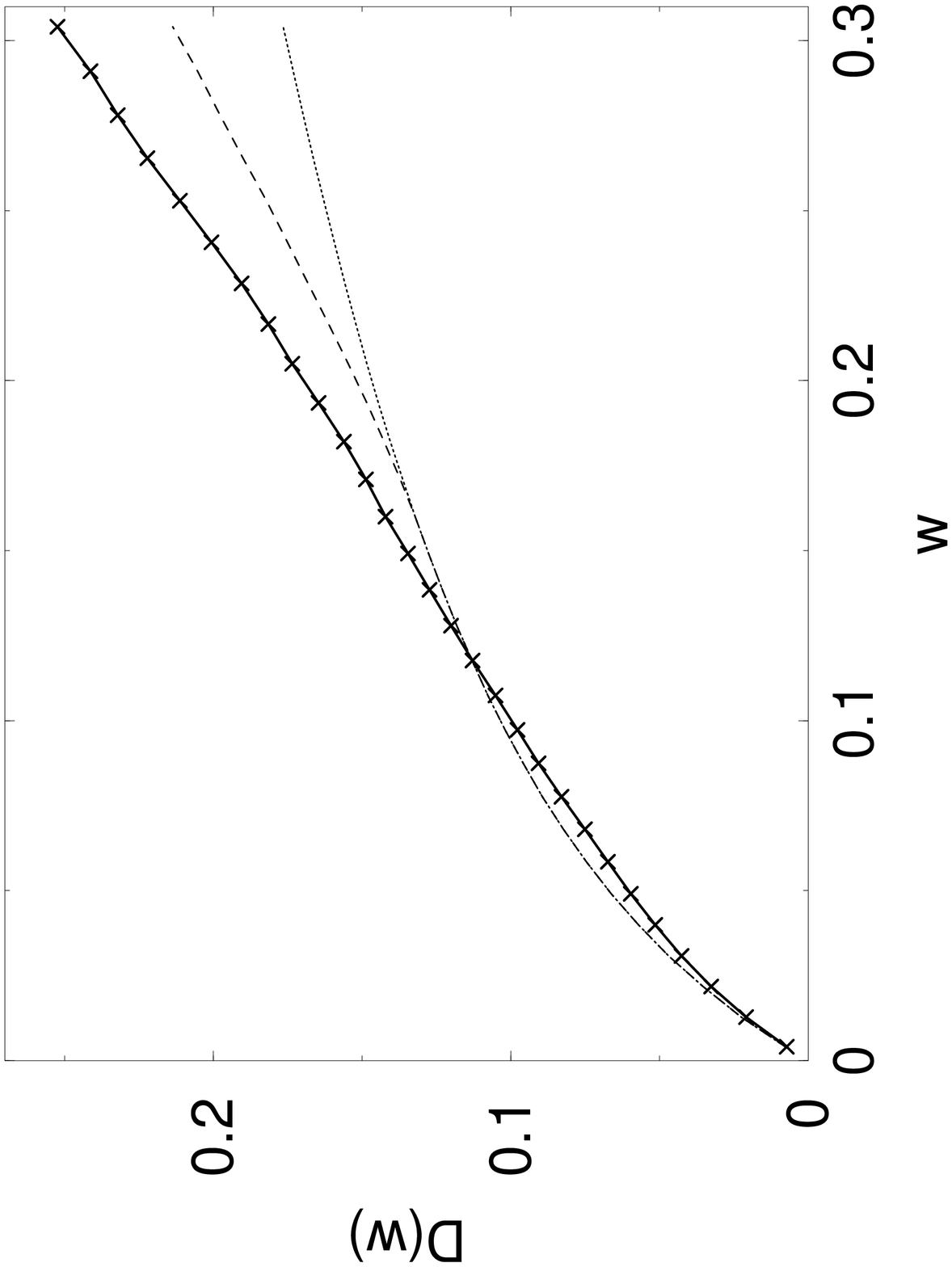}\hspace*{-0.31cm}}}
\end{center}
\caption{Correction of the Machta-Zwanzig approximation by
collisionless flights: (a) probability $p_{\rm cf}$ of collisionless
flights across a trap, numerical results (dotted line with crosses),
and the analytical approximation Eq.\ (\ref{eq:pcfa}) (bold line) (b)
diffusion coefficient $D$, numerical results (bold line with
crosses) in comparison to the Machta-Zwanzig approximation Eq.\
(\ref{eq:mz}) (dotted line), and compared to the correction of Eq.\
(\ref{eq:cf}) by including $p_{\rm cf}$ (dashed line).}
\label{fig3}
\end{figure}

As Machta and Zwanzig remarked themselves, for larger
$w$ there is a non-vanishing probability $p_{\rm cf}$ for the particle
to move collision-free across a trap. In Fig.\ \ref{fig3} (a), we have
calculated this probability from computer simulations as well as in a
straightforward analytical approximation, which is described in Appendix
\ref{appa}. Note that collisionless flights occur only for
$w>0.1547$, see Eq.\ (\ref{eq:w0}).

If we rely on the Machta-Zwanzig picture of diffusion as a hopping
process with frequency $\tau^{-1}$ over distances $l$, we can correct
Eq.\ (\ref{eq:mz}) in the following way: If a particle moves
collision-free across a trap, it travels within the time $\tau$ over a
{\em larger} distance than assumed in Eq.\ (\ref{eq:mz}).  For this
larger distance we take the distance between a center of a trap and
the center of its next nearest neighbor which is
$l_2=\sqrt{3}l$. These processes would thus yield a larger diffusion
coefficient $D_{l_2}=l_2^2/(4\tau)=3D_{MZ}$, where $D_{MZ}$ is the
diffusion coefficient of Eq.\ (\ref{eq:mz}). We now define the
corrected Machta-Zwanzig diffusion coefficient $D_{\rm cf}$ by
weighting the contribution of collisionless flights via the
probability $p_{\rm cf}$ of Fig.\ \ref{fig3} (a),
\begin{eqnarray}
D_{\rm cf}&=&[1-p_{\rm cf}]D_{MZ}+p_{\rm cf}3D_{MZ} \nonumber \\
 &=&[1+2p_{\rm cf}]D_{MZ}\quad . \label{eq:cf}
\end{eqnarray}
The result is plotted in Fig.\ \ref{fig3} (b). For $w>0.1547$, the
revised formula Eq.\ (\ref{eq:cf}) improves the original
Machta-Zwanzig approximation considerably, however, it is still much
smaller than the numerically exact results.

\subsection{Probability of backscattering}

\begin{figure}
\begin{center}
\epsfxsize=5.5cm
\subfigure[\hspace*{4.4cm}]{\rotate[r]{\epsfbox{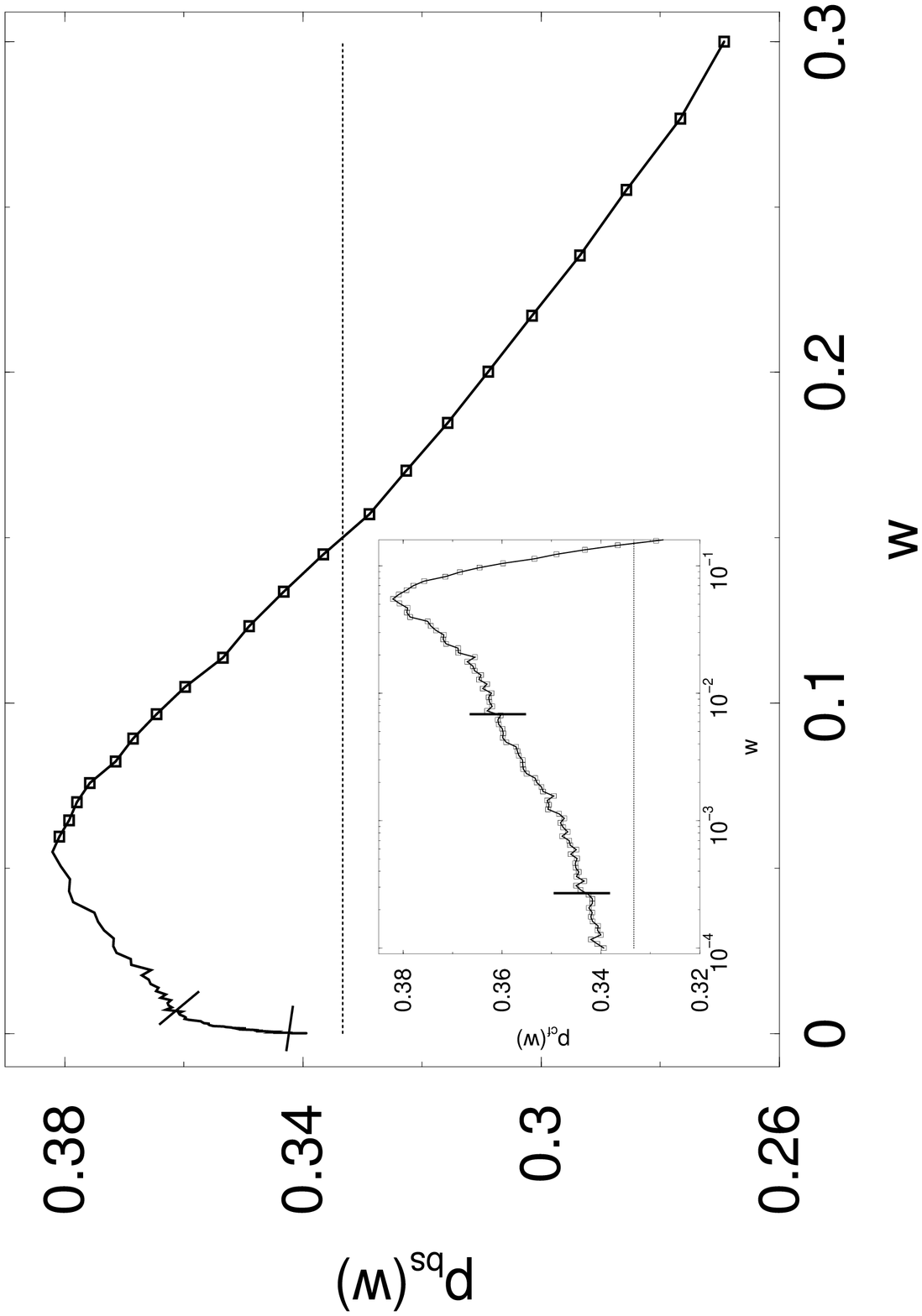}\hspace*{-0.5cm}}}
\hspace*{0.6cm}
\epsfxsize=5.5cm
\subfigure[\hspace*{5.5cm}]{\rotate[r]{\epsfbox{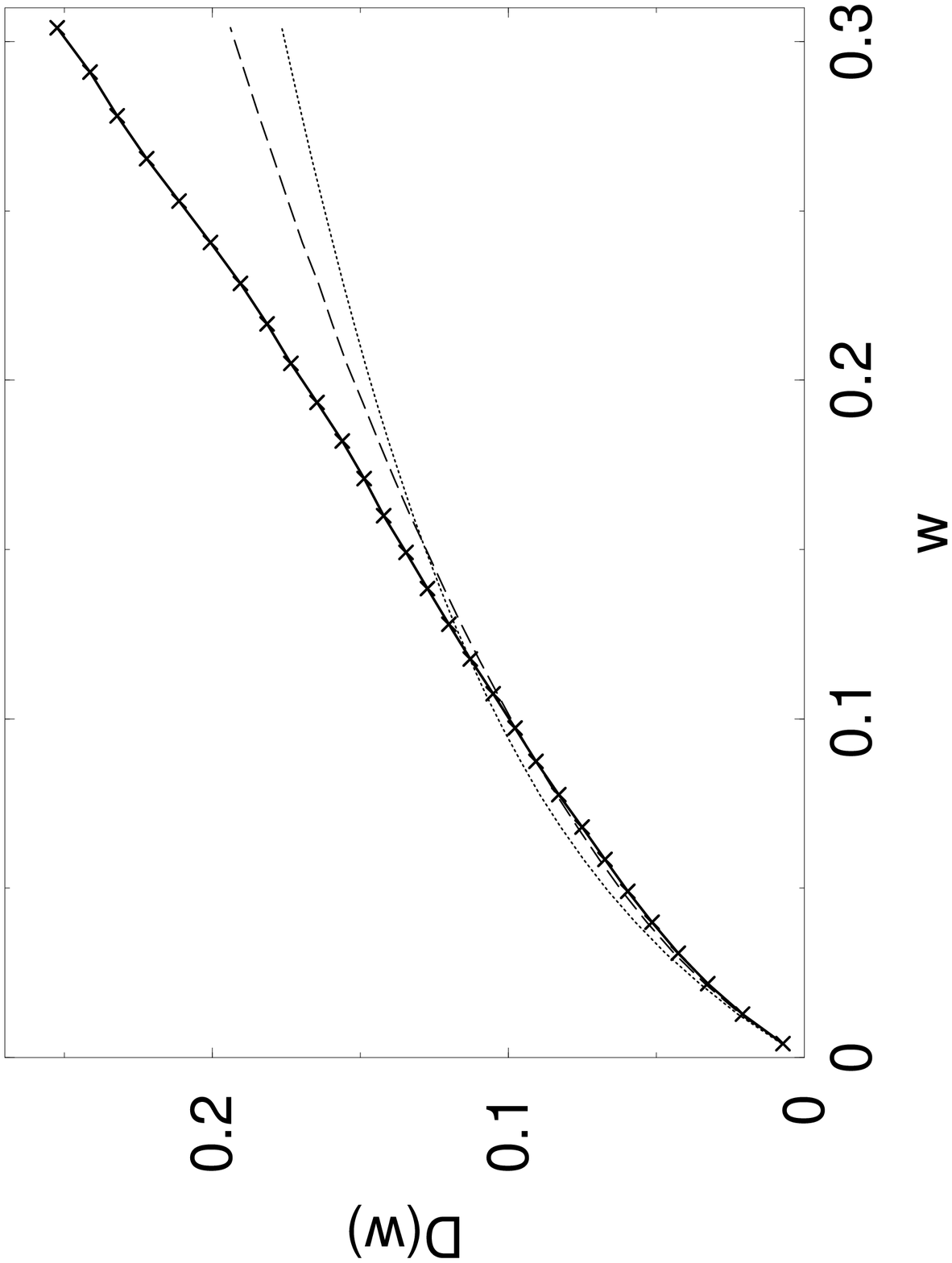}\hspace*{-0.16cm}}}
\end{center}
\caption{Correction of the Machta-Zwanzig approximation by
backscattering: (a) backscattering probability $p_{\rm bs}$ as a
function of the gap size $w$. For larger $w$ the single data points
are plotted by symbols and are connected with lines, for smaller $w$
only the lines are shown. The dotted line corresponds to the value of
$1/3$ of equal probability for {\em any} gap, as it is assumed in the
Machta-Zwanzig approximation. The inset is a half-logarithmic blowup
of the initial region for small $w$. The bars included in the figure
refer approximately to the regions of different slope in the main
figure. (b) numerical results for the diffusion coefficient (bold line
with crosses) in comparison to the Machta-Zwanzig approximation Eq.\
(\ref{eq:mz}) (dotted line), and to the correction via $p_{\rm bs}$
Eq.\ (\ref{eq:cf}) (dashed line).}
\label{fig4}
\end{figure}

A second significant contribution to the correction of Eq.\
(\ref{eq:mz}) is determined by the backscattering probability $p_{\rm
bs}$, which is the probability of the moving particle to leave the
trap through the same gap where it entered. We computed $p_{\rm bs}$
numerically by repeatedly injecting the particle through a specific
gap and observing trough which gap it left the trap. The particles are
initially situated at one entrance of a trap, and they are uniformly
distributed in the respective phase space of this entrance, which
consists of the position of a particle on the entrance line, $-1\ge
2x/w \ge 1$, and of the sine of the angle between the velocity
direction and an axis perpendicular to the entrance, $-1\ge
\sin \alpha \ge 1$. Alternatively, $p_{\rm bs}$ can be determined from
a single long trajectory. The backscattering probability obtained from
our simulations is shown in Fig.\ \ref{fig4} (a).

The Markovian approximation of Machta and Zwanzig Eq.\ (\ref{eq:mz})
corresponds to a backscattering probability of $1/3$. However, for
small values of $w$ the numerically computed probability is
significantly larger than $1/3$, whereas for larger $w$ it is
considerably smaller. The reduced probability for backward scattering
at larger $w$ is in part due to collisionless flights across the
trap. It is remarkable that the backscattering probability is
different from $1/3$ even for gap sizes where the average number of
collisions between inter-trap hops is large. At $w=0.055$, where the
backscattering probability reaches its maximum value of $p_{\rm
bs}=0.38$, more than 17 collisions occur before the particle hops to
the next trap and for $w=0.02$, where $p_{\rm bs}$ is still close to
$0.36$, the number of collisions is greater than
$50$. Correspondingly, the detailed functional form of $p_{\rm bs}$
appears to be quite intricate: Below the maximum at $w=0.055$ there
are at least three regions where $p_{\rm bs}$ decreases approximately
linear in $w$ with different values of the slope. As the numerical
results indicate, the function must eventually drop extremely sharply
to $p_{\rm bs}(0)=1/3$. Details of these regions are shown in the
magnification included in the figure. The vertical lines separating
different regions correspond to the respective lines separating
regions of different slope in the main figure. A very close look
reveals that the fine structure of all these different regions appears
to be quite similar, however, we note that this structure is
essentially within the range of our numerical errors.

\begin{figure}
\epsfxsize=8cm
\centerline{{\epsfbox{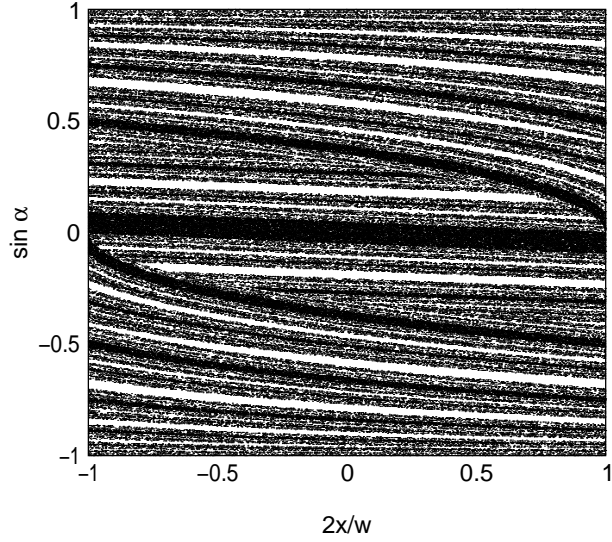}}}
\vspace*{0.3cm}
\caption{Initial conditions yielding backscattering for $w=0.1$. Each
dot represents an initial condition for which the particle leaves the
trap through the same aperture where it entered. $2x/w$ is the
position of a particle on the line of the entrance, and $\alpha$ is
the angle of flight of an in-going particle to an axis perpendicular
to the entrance. }
\label{fig5}
\end{figure}

Fig.\ \ref{fig5} shows the initial conditions in the $(x,\sin
\alpha)$-plane leading to backscattering for a gap size of
$w=0.1$. Here, the $x$-axis is parallel to the trap entrance and the
origin is located at the center of the entrance.  Initial conditions
were drawn at random from a uniform distribution in this plane and a
dot was plotted whenever the particle left the trap through the same
aperture where it entered. Simulations at different densities yield
similar pictures. The complicated, intertwined structure is
reminiscent of a fractal set. The modifications of this structure by
varying $w$ must be related to the changes seen in the function
$p_{\rm bs}$ of Fig.\ \ref{fig4} (a). This may explain why, even on
the coarse scale of \ref{fig4} (a), $p_{\rm bs}$ is not a simple
function of $w$. Moreover, the detailed changes of this fractal set
may be reflected in the respective detailed changes of $p_{\rm bs}$ on
the fine scale, as illustrated in the magnification.

Having the probability of backscattering $p_{\rm bs}$ we can now
perform a second correction of the original Machta-Zwanzig diffusion
coefficient Eq.\ (\ref{eq:mz}). For this purpose, we again assume that
diffusion can be treated as a hopping process with a frequency
$\tau^{-1}$ over distances $l$. We now inquire to which traps
the particle can move by performing two jumps within a total time
interval of $2\tau$. There are only two possibilities: either the
particle suffers backscattering, that is, it goes back to its original
trap and does not contribute to any actual displacement within
$2\tau$, or it moves over a distance $l_2$ to the left or to the
right of its original trap. Thus, the corresponding diffusion
coefficient reads
\begin{eqnarray}
D_{\rm bs}&=&\frac{[1-p_{\rm bs}]l_2^2}{8\tau} \nonumber \\
&=&[1-p_{\rm bs}]3/2D_{MZ}\quad . \label{eq:bs}
\end{eqnarray}
$D_{\rm bs}$ is plotted in comparison to our Einstein formula results,
and to the original Machta-Zwanzig Eq.\ (\ref{eq:mz}), in Fig.\
\ref{fig4} (b). For smaller $w$, the corrected diffusion coefficient
approximates the numerically exact values quite well. Thus, we
conclude that the existence of backscattering is basically responsible
for the Machta-Zwanzig argument overestimating diffusion for small
values of $w$. For larger $w$ Eq.\ (\ref{eq:bs}) again improves the
original Machta-Zwanzig approximation, like the previous approximation
Eq.\ (\ref{eq:cf}), however, like this it yields much smaller results
than the correct values.

\subsection{Combining collisionless flights, backscattering, and
a symbolic dynamics}\label{hoc}

So far we have identified two different microscopic scattering
mechanisms which were not included in the original Machta-Zwanzig
approximation for the diffusion coefficient, and which evidently play
a significant role for understanding the full density-dependent
diffusion coefficient: Namely, the probability of collisionless
flights across a trap leading to the diffusion coefficient of Eq.\
(\ref{eq:cf}), and the probability of backscattering leading to Eq.\
(\ref{eq:bs}). Consequently, one may think of combining these two
different approximations within a single diffusion coefficient
formula. This can be performed by simply replacing the Machta-Zwanzig
diffusion coefficient $D_{MZ}$ in Eq.\ (\ref{eq:bs}) by $D_{\rm cf}$
of Eq.\ (\ref{eq:cf}) yielding
\begin{equation}
D_{1}=3/2[1-p_{\rm bs}][1+2p_{\rm cf}]D_{MZ} \quad . \label{eq:comb}
\end{equation}
This function, denoted as a first order approximation, is shown in
Fig.\ \ref{fig6} (a) in comparison to the numerically exact results
and to the Machta-Zwanzig Eq.\ (\ref{eq:mz}). For larger $w$, this
combined approximation is much closer to the numerical results than
the original Machta-Zwanzig formulation, however, there still remains
a notable quantitative difference.

In the same figure, corrections of higher order based on the idea of
Eq.\ (\ref{eq:comb}) are depicted. In the following we just outline 
the basic concept, the detailed formulas are then given in Appendix
\ref{appb}. The higher-order corrections of the diffusion coefficient
are obtained by numerically computing the probabilities of
higher-order backscattering, and by building them into a respectively
generalized version of Eq.\ (\ref{eq:bs}). This generalized expression
of the backscattering diffusion coefficient $D_{\rm bs}$ is then
combined with the respective collisionless flight-diffusion
coefficient $D_{\rm cf}$ of Eq.\ (\ref{eq:cf}) in the same way as
before. The probabilities of higher-order backscattering have been
computed on the basis of a simple symbolic dynamics, as it can be
defined in case of simple backscattering: We followed a long
trajectory of a particle in the Lorentz gas. For each visited trap we
labeled the entrances through which the particle entered with $z$, the
exit to the left of this entrance with $l$, and the one to the right
with $r$. Thus, a trajectory in the Lorentz gas can be mapped to a
sequence of symbols $z, l$, and $r$. Note that the symbolic dynamics
we are using here is different to the one applied in other work in
that we are labeling the three gaps of a trap, whereas in previous
work the single disks accessible after a collision have been chosen,
which required an alphabet of 12 symbols
\cite{Cvit95,Gasp}. $p(z)=p_{\rm bs}$ is then the backscattering
probability depicted in Fig.\ \ref{fig4} (a), whereas, due to
symmetry, $p(l)=p(r)=(1-p(z))/2$ corresponds to forward
scattering. Combining these probabilities with the correction by
collisionless flights yields Eq.\ (\ref{eq:comb}) as a first order
approximation.

Higher-order correlations can be calculated by taking into account the
probabilities of longer symbol sequences.  For example, the second
order approximation involves probabilities corresponding to nine
symbol sequences each consisting of two symbols, $p(zz)$, $p(zl)$,
$p(zr)$, $p(lz)$, $p(ll)$, $p(lr)$, $p(rz)$, $p(rl)$, $p(rr)$. In
complete analogy to Eq.\ (\ref{eq:comb}), a respective second-order
approximation of the diffusion coefficient is then computed on the
basis of these numerical probabilities, and by associating to them the
respective distances traveled within a time interval of $3\tau$, see
Eq.\ (\ref{eq:d2}). In the third-order approximation the probabilities
correspond to sequences of three symbols, for example, three times
backscattering within a time interval of $4\tau$ corresponding to
$p(zzz)$, etc., and lead to the diffusion coefficient of Eq.\
(\ref{eq:d3}). In general, the $k$-th order approximation involves the
probability of sequences of $k$ symbols. Fig.\ref{fig6} (a) shows that
by including such correlations the respective higher-order
approximations of the diffusion coefficient converge, and basically
move closer to the numerical exact results. However, for $w>0.1$ the
approximations converge to the exact results apparently only globally,
and not locally. A clear sign of this is that in third order the
respective approximation already exceeds the numerical data for large
$w$, although the approximations seem to approach the exact data
generically from below. On the other hand, one should take into
account that this approximation scheme is based on a purely
heuristical ansatz for the diffusion coefficient. Thus, an exact
convergence should not necessarily be expected.

We therefore studied the effect of increased (or decreased)
backscattering with the help of a lattice gas computer simulation. In
such a simulation the Lorentz gas is mapped to a honeycomb lattice
where the sites of the lattice represent the traps. The moving
particle hops from site to site with frequency $\tau^{-1}$ , which is
identical to the exact hopping frequency used in Machta-Zwanzig
theory. We first describe the first order approximation: At each step
the particle hops back to the site where it came from with probability
$p_{\rm bs}$ or to one of the other sites with probability $(1-p_{\rm
bs})/2$. The backscattering probability $p_{\rm bs}$ used in the
lattice gas simulations is the one numerically obtained from
simulations in the Lorentz gas. Also in the lattice gas simulations we
determine the diffusion coefficient from the Einstein formula Eq.\
(\ref{eq:einst}). The results of these simulations are shown in Fig.\
\ref{fig6} (b) by a dotted line. Taking into account these first order
correlations brings the diffusion coefficient closer to the correct
value, but there still is a considerable deviation. This indicates the
importance of higher order correlations.

These higher order correlations can be obtained by correlating more
than two hops between neighboring traps. To determine the diffusion
coefficient for such multiple hopping events by lattice gas
simulations we used the probabilities $p(lrz)$ calculated up to fourth
order from long trajectories in the Lorentz gas. The results of these
calculations are shown in Fig.\ \ref{fig6} (b). As can be seen in the
figure, the diffusion coefficient obtained from this scheme converges
very quickly to the numerically exact results, in particular for small
$w$. For larger $w$, the convergence is somewhat slower, however, the
fourth order approximation can be hardly distinguished from the
numerically exact results on the scale of Fig.\ \ref{fig6} (b). For
growing order the diffusion coefficient converges to the correct
results exactly.

\section{Boltzmann approximation for the diffusion coefficient}

In the previous section, we have discussed how the Machta-Zwanzig theory
can be improved systematically. An alternative approach for understanding 
diffusion in the Lorentz gas may be obtained from
kinetic theory, which essentially applies at low densities. In Refs.\
\cite{WvL,Bruin}, the Boltzmann approximation for the random Lorentz gas,
where the scatterers are distributed randomly in the plane without
overlap, has been used to compute the diffusion coefficient to
\begin{equation}
D_{\rm Bo}(n)=\frac{3}{8}l_c(n) v \quad , \label{eq:dbr}
\end{equation}
where $l_c=1/(2Rn)$ is the collision length of the moving particle.
At higher number densities $n$ the excluded volume of the scatterers
becomes important, and the respective shorter collision length reads
\begin{equation}
D_{\rm Bo}(n)=\frac{3}{16}(\frac{1}{n}-\pi) \quad , \label{eq:dbp}
\end{equation}
where we have changed to units with $v=1$ and $R=1$. The same result
for the collision length can be obtained directly from the
Machta-Zwanzig argument of Section II \cite{MaZw83}. In case of the
periodic Lorentz gas $n$ is related to the gap size $w$ by Eq.\
(\ref{eq:nd}). Note that for $w=0$, where the particle is trapped, the
corresponding density is $n<\infty$ and the collision length is still
finite.  Thus, Eq.\ (\ref{eq:dbp}) implies a diffusion coefficient of
$D_{\rm Bo}(w=0)>0$, that is, this approximation inevitably leads to
an offset between $D_{\rm Bo}(w=0)$ and the exact result of
$D(w=0)=0$. Therefore, it can immediately be concluded that $D_{\rm
Bo}$ is not correct at the highest densities.

In Fig.\ \ref{fig7}, Eq.\ (\ref{eq:dbp}) is shown in comparison to the
numerically exact results and to the Machta-Zwanzig approximation Eq.\
(\ref{eq:mz}). For this purpose, the diffusion coefficient has been
plotted as a function of $1/n$, and an offset of $\Delta D=0.025$ has
been subtracted from Eq.\ (\ref{eq:dbp}). According to this equation,
$D_{\rm Bo}(n)$ should go linear in $1/n$ with a slope of $3/16$,
and the offset $\Delta D$ is the only fit parameter. Indeed, Fig.\
\ref{fig7} shows that, on a sufficiently coarse scale and over a wide
range of lower densities, $D$ matches quite well to this functional
form. In particular, the value of the slope appears to be almost
exact. Only for larger densities $D$ deviates from linearity in
$1/n$. However, here the Machta-Zwanzig approximation seems to
describe the functional form at least qualitatively quite well.

From a conceptual point of view one may inquire why a Boltzmann
description should at all be applicable to the periodic Lorentz
gas. The basic difficulty is that because of the occurrence of an
infinite horizon, a diffusion coefficient exists in the periodic
Lorentz gas only in the regime of very high densities, where a simple
Boltzmann approximation may not necessarily be expected to work. We
first remark that in a Lorentz gas there are no density corrections
due to screening of particles, or to many-particle collisions, as they
are contained in an Enskog equation \cite{ChCo}. However, the
existence of an offset at large densities may be related to the fact
that the Boltzmann approximation represents only the first term in a
series expansion in $n$, which is known as the density expansion of
kinetic theory. For the random Lorentz gas, such a density expansion
has been carried out explicitly in Refs.\ \cite{WvL,Bruin}. It has
been found that there exist higher-order terms being logarithmic in
the density corresponding to so-called ring collisions, which diminish
the Boltzmann diffusion coefficient quantitatively. However, these
logarithmic corrections are difficult to see in the functional form of
the diffusion coefficient. It is not known to us whether a density
expansion has been performed for the periodic Lorentz gas. But the
offset we find might be associated to the existence of such
higher-order corrections in the density, and the apparent
Boltzmann-like linear behavior may be related to the observation that
in similar systems higher-order corrections do not change the
functional form of the diffusion coefficient in a drastic way.

Because of the functional relation between number
density $n$ and gap size $w$ Eq.\ (\ref{eq:nd}), and according to the
Boltzmann approximation Eq.\ (\ref{eq:dbp}), there should exist a
respective approximate linear dependence of the diffusion coefficient
$D$ on $w$, as can in fact be seen in Fig.\
\ref{fig2}. The quadratic term in $w$ which appears in Eq.\
(\ref{eq:nd}) turns out to be quantitatively at most $6\%$ for largest
$w$, and, after a proper adjustment, cannot be seen as any deviation
to a linearity in $w$.

\section{Fine structure of the diffusion coefficient}

To learn more about the very detailed dependence of $D$ on the density
$n$, or alternatively on the gap size $w$, we performed further
computer simulations in the region of large $w$. The results are shown
in Fig.\ \ref{fig8}. Here, the respective residuals of $D$ have been
plotted, that is, the deviation of the diffusion coefficients from a
linear fit in $w$ over the whole region shown in the figure. At each
value of $w$ ten independent runs of more than $2\times 10^9$
collisions each have been carried out yielding the diffusion
coefficients shown by the dots. The squares correspond to the averages
over the ten runs, where the size of the squares indicates the size of
the numerical error.  Fig.\ \ref{fig8} may be regarded as a signature
of long-range correlations in microscopic scattering events. A certain
subclass of such scattering events are the ring collisions mentioned
above leading to a series of logarithmic corrections, and divergences,
in the density expansion of the diffusion coefficient. It might be
conjectured that the existence of these divergences in the density
expansion of kinetic theory is due to the problem of approximating a
diffusion coefficient, which is in fact a function as complicated as
the one shown in Fig.\ \ref{fig8}, in form of a simple series
expansion, as has already been remarked in Ref.\ \cite{Gasp}.

\section{Conclusions}

We have numerically computed the density dependence of the diffusion
coefficient in the two-dimensional periodic Lorentz gas. A comparison
of our results with the analytical approximation of Machta and
Zwanzig shows that their argument is only
asymptotically correct in the limit of high densities. We have
furthermore discussed how their approximation can be  
corrected systematically by including correlations in the hopping mechanism.

As an alternative way to understand the structure of the
density-dependent diffusion coefficient, we compared our numerical
results to a simple Boltzmann approximation. We found that on a coarse
scale there exists a dynamical crossover from a function linear in
$1/n$ for lower densities, as qualitatively be described by the
Boltzmann diffusion coefficient, to a behavior at high densities which
can qualitatively be understood by the Machta-Zwanzig argument. 
Our two different
points of view of understanding the density-dependent diffusion
coefficient - namely, on the basis of the Machta-Zwanzig argument, and
by a Boltzmann approximation -, are not at all contradictory. The
complicated microscopic scattering processes responsible for the
failure of the Machta-Zwanzig approximation apparently just superpose
in a way such that the resulting diffusion coefficient is approximately
linear in $1/n$ on a coarse scale. A similar relation between
nonlinear corrections and a linear response has been discussed in
Refs.\ \cite{CoRo98,Ch2} with respect to the existence of Ohm's law in
a periodic Lorentz gas with an external field. That this picture is 
too simple is demonstrated by the fact that on a very
fine scale we find deviations from such a simple linear behavior.
These tiny fluctuations may be regarded as the signature of specific microscopic
correlated scattering processes in the deterministic diffusion
coefficient.

A dynamical crossover between different
asymptotic laws for a parameter-dependent diffusion coefficient has
already been found and discussed in related periodic one-dimensional
chaotic maps \cite{RKdiss,dcrc}. As in our discussion above, for these
maps two different random walk models have been used to describe the
corresponding asymptotic behavior: For small parameters, the respective
random walk basically depends on the hopping probability of a particle
leaving a cell, thus corresponding to the Machta-Zwanzig
approximation, whereas for larger parameters the respective random
walk scales with the distance a particle travels per time step, thus
corresponding to the Boltzmann approximation, where diffusion is
proportional to the collision length of the moving particle. We
furthermore note that the diffusion coefficients in this kind of maps
have been found to be fractal with respect to variation of the control
parameter, and that specific correlated microscopic scattering events
could be identified as being responsible for this fractal structure
\cite{RKD,RKdiss}. The same phenomena have been encountered in related
two-dimensional multi-baker maps, which are simple toy models of the
periodic Lorentz gas \cite{GaKl}.

Whether the density-dependent diffusion coefficient of the periodic
Lorentz gas is in fact a non-differentiable function of the parameter,
as it is the case for the one-dimensional maps discussed above,
remains an open problem. It seems to us that this question cannot be
answered by performing straightforward computer simulations, but that,
instead, more refined techniques to compute the diffusion coefficient
are necessary, as they have been developed, for example, in Ref.\
\cite{RKdiss} in case of one-dimensional maps. A solution to this
problem appears to be important for a complete understanding of
deterministic diffusion in the periodic Lorentz gas.\\

\noindent {\bf Acknowledgments}\\
This article is dedicated to G. Nicolis on occasion of his 60th
birthday. R.K. wants to express his gratitude to G. Nicolis for many
scientific discussions and advice, and for his continuing support over
the last two years. In addition, R. K. is indebted to P. Gaspard for
his interest in this problem, and for many valuable
discussions. Helpful comments by E. Barkai, H. van Beijeren and
D. Panja are gratefully acknowledged. Finally, R.K. thanks the
European Commission for financial support within its Training and
Mobility Program.

\begin{appendix}
\section{Probability for a collisionless flight across a
trap}\label{appa} 

In this part of the Appendix we describe a simple analytical
approximation by which the probability $p_{\rm cf}$ for a collisionless
flight across a trap can be computed. The result is shown in Fig.\
\ref{fig3} (a). Our approximation is based on straightforward applying
the Machta-Zwanzig phase space argument described in Section II. It
says that
\begin{equation}
p_{\rm cf}=\frac{S v \tau \phi}{2\pi A} \quad , \label{eq:pcf}
\end{equation}
where $A$ is the area of the trap, $2\pi$ is the measure of the
velocity space of the trap, $S$ is the region of the boundary of the
trap across which collisionless flights are possible, $v=1$ is the
velocity of the moving particle, $\tau$ is the average time at which a
particle leaves the trap, and
\begin{equation}
\phi=\int d\gamma\: \underline{n}\cdot\underline{v}\: \varrho(\gamma) \label{eq:phi}
\end{equation}
is the average flux of particles with velocity $\underline{v}$ parallel to
the normal $\underline{n}$ at one entrance, with $\gamma$ being the angle
between $\underline{n}$ and $\underline{v}$, and $\varrho(\gamma)=\cos\gamma$
being the invariant probability distribution of $\gamma$ in
equilibrium. Here we have already assumed that, in an approximation,
$\phi$ is completely independent of its position at the boundary. In
Ref.\ \cite{MaZw83} it has been noted that
\begin{equation}
\tau=\frac{\pi A}{3w} \quad ,
\end{equation}
where $w$ is the size of an entrance, such that Eq.\ (\ref{eq:pcf}) reads 
\begin{equation}
p_{\rm cf}=\frac{S\phi}{6w} \quad .
\end{equation}
Thus, it remains to compute $S$ and $\phi$. The relevant variables for
this purpose, and the geometry, are defined in Fig.\ \ref{fig9},
where, because of symmetry, we have reduced the problem to one
entrance. In a second approximation, we assume that regions of size
$e$, as the one indicated in the figure, are the only parts of the
boundary which contribute to free flights. $e$ is precisely defined as
the section on the entrance line between the two horizontal lines
which touch the upper scatterer tangentially, and which go from the
lower edge of the left entrance to the lower edge of the right
entrance, respectively. Because of symmetry, there are six regions
like this for each trap leading to $S=6e$. $e$ can simply be
calculated to
\begin{equation}
e=\frac{\sqrt{3}(2+w)}{2\sin(\pi/3)}-\frac{1}{\sin(\pi/3)}-1 \quad .
\label{eq:e} 
\end{equation}
Note that only for $e>0$ collisionless flights across a trap are
possible. In other words, $e=0$ or, by employing Eq.\ (\ref{eq:e}),
\begin{equation}
w_0=\frac{2(\sin(\pi/3)+1)}{\sqrt{3}}-2\simeq0.1547 \label{eq:w0}
\end{equation}
marks just the onset of the existence of collisionless flights with
respect to varying $w$.

It now remains to compute the flux $\phi$ in these regions. To do
this, we need to know the limits of the integral in Eq.\
(\ref{eq:phi}). As a very crude approximation, we assume that these
boundaries are determined by the angles $\beta$ and $\delta$ such that
\begin{eqnarray}
\phi&=&\int_{\beta-\delta}^{\beta+\delta}d\gamma\: \cos^2\gamma
\nonumber \\
&=&\delta+1/2\cos(\pi/12)\sin(2\delta) \quad . \label{eq:phid}
\end{eqnarray}
Here, $\beta=\pi/6$ is identical to the angle between a horizontal
velocity vector $\underline{v}$ and the normal $\underline{n}$ of the entrance.
$\delta$ gives a bound for the directions the velocity $\underline{v}$ can
take such that a particle going in through the region of size $e$ can
perform a collisionless flight across the trap. By a straightforward
calculation, it can be found that
\begin{equation}
\delta=\arctan\frac{2e\cos(\pi/6)}{2+w-2(e+1)\cos(\pi/3)+e\sin(\pi/6)}
\quad . \label{eq:d}
\end{equation}
In summary, we have computed the probability $p_{\rm cf}$ for
collisionless flights across a trap as
\begin{equation}
p_{\rm cf}=\frac{e\phi}{w} \quad , \label{eq:pcfa}
\end{equation}
where $e$ is given by Eq.\ (\ref{eq:e}), and $\phi$ is given by Eqs.\
(\ref{eq:phid}) and (\ref{eq:d}). 

\section{Higher-order backscattering approximations of the
diffusion coefficient}\label{appb}

Here we give the explicit formulas for higher-order approximations of
the diffusion coefficient in the spirit of Eq.\ (\ref{eq:comb}), where
this equation represents already the first order correction. The basic
idea of these corrections is discussed in Section \ref{hoc}.

The underlying picture for all these approximations is the
Machta-Zwanzig assumption of diffusion as a hopping process with
frequency $\tau^{-1}$ on a hexagonal lattice of traps, where the
lattice sites are separated by a distance of $l$. We use in the
following the symbolic dynamics introduced in Section \ref{hoc} to
denote the transition probabilities of a particle which hops from one
initial trap to some other trap on the lattice within a certain time
interval of multiples of $\tau$. For the particular time interval of
$3\tau$, we find that the particle can move\\

\noindent -- with a probability $p_1=p(zz)+p(zl)+p(zr)+p(lz)+p(rz)$
over a distance $l_1=l=(2+w)\sqrt{3}$\\
-- with a probability $p_3=p(lr)+p(rl)$ over a distance
$l_3=\sqrt{7}l$\\
-- with a probability $p_{3'}=p(ll)+p(rr)$ over a distance
$l_{3'}=2l$.\\

The corresponding second-order diffusion coefficient then reads
\begin{equation}
D_{2}=(12\tau)^{-1}[p_1l_1^2+p_3l_3^2+p_{3'}l_{3'}^2] \quad
. \label{eq:d2} 
\end{equation}
In the same way the third-order approximation is obtained. Here, we
have a collection of 27 probabilities corresponding to the different
lattice sites where the particle can move within a time interval of
$4\tau$. These transition probabilities are labeled by all possible
combinations of the three symbols l, r, and z corresponding to
sequences of three symbols. Thus, the particle can move\\

\noindent -- with a probability $p_0=p(lzz)+p(rzz)+p(zlz)+p(zrz)+p(zzz)$ 
over a distance $l_0=0$\\
-- with a probability
$p_2=p(lll)+p(llz)+p(lrz)+p(lzl)+p(lzr)+p(rlz)+p(rrr)+p(rrz)+p(rzl)+p(rzr)+p(zll)+p(zlr)+p(zrl)+p(zrr)+p(zzl)+p(zzr)$ 
over a distance $l_2=\sqrt{3}l$\\
-- with a probability $p_{41}=p(lrl)+p(rlr)$ over a distance
$l_{41}=3l$\\
-- with a probability $p_{42}=p(llr)+pp(lrr)+p(rll)+p(rrl)$ over a
distance $l_{42}=(1.5+\sqrt{3})l$.\\

\begin{figure}
\begin{center}
\epsfxsize=5.5cm
\subfigure[\hspace*{5.4cm}]{\rotate[r]{\epsfbox{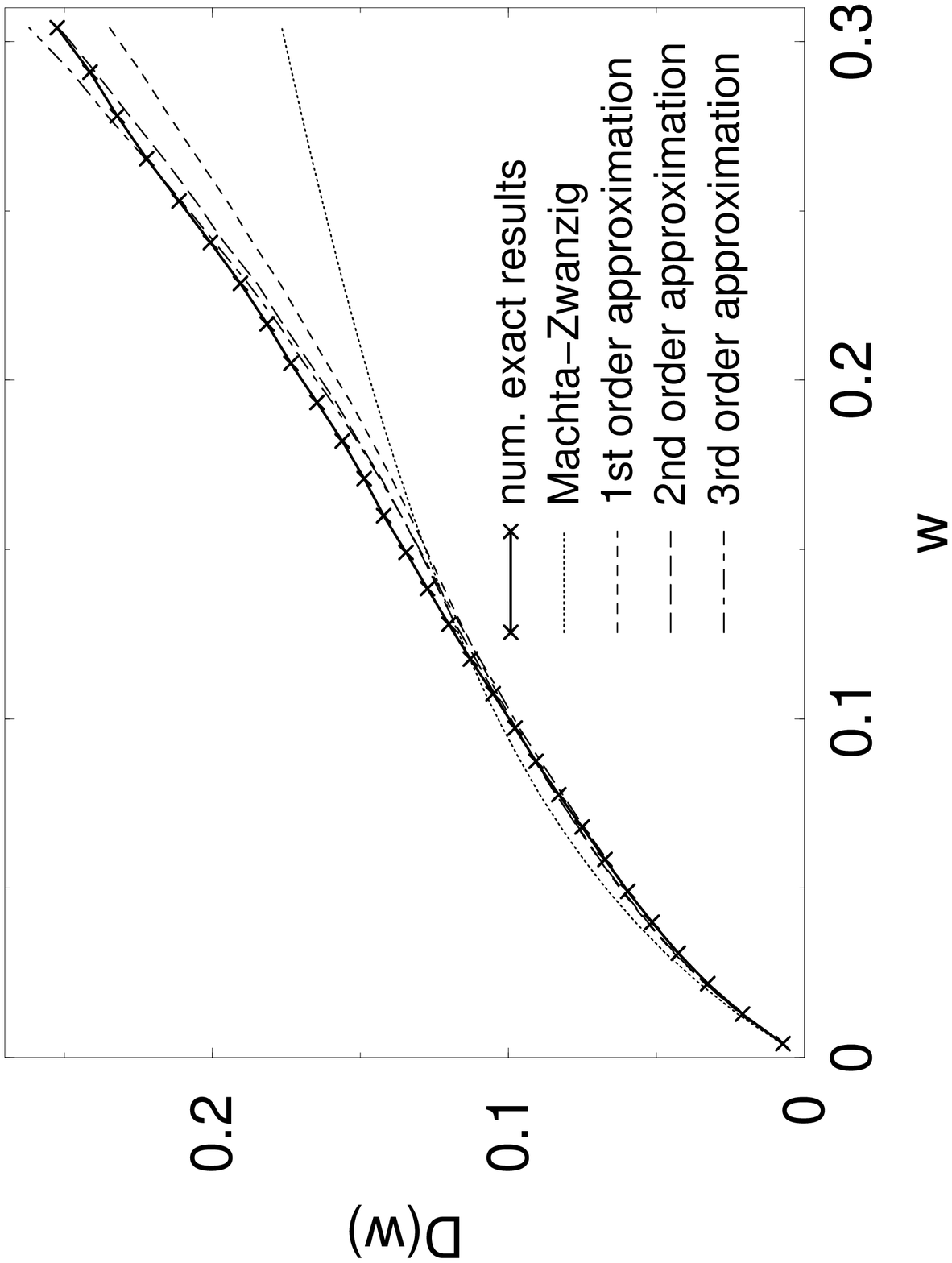}\hspace*{-0cm}}}
\hspace*{0.3cm}
\epsfxsize=5.5cm
\subfigure[\hspace*{5.4cm}]{\rotate[r]{\epsfbox{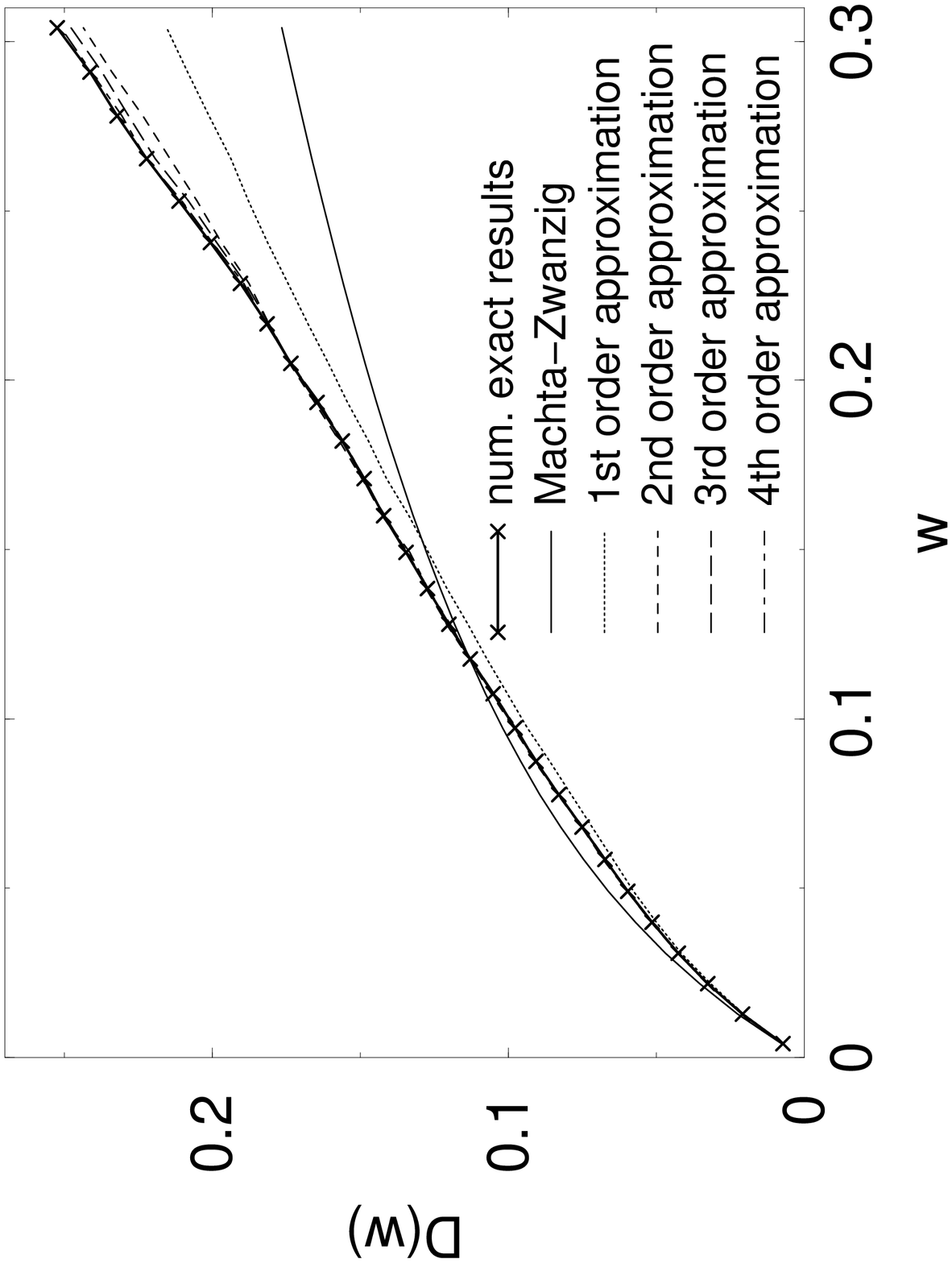}\hspace*{-0cm}}}
\end{center}
\caption{Systematic approximations of the exact diffusion coefficient
in higher order: (a) combined approach of employing the numerical
probabilities of collisionless flights and of backscattering within a
hierarchy of analytical expressions of the diffusion coefficient. The
first-order approximation is given by Eq.\ (\ref{eq:comb}) by using
the probabilities $p_{\rm cf}$ and $p_{\rm bs}$ of Figs.\ \ref{fig3}
(a) and \ref{fig4} (a). The second order corresponds to Eq.\
(\ref{eq:d2}), and the third order corresponds to Eq.\
(\ref{eq:d3}). They include higher-order backscattering events, as
described in the text and in Appendix \ref{appb}. (b) diffusion
coefficients obtained from lattice gas simulation by using
probabilities of correlated hopping sequences as calculated in Lorentz
gas simulations (see text).}
\label{fig6}
\end{figure}

\begin{figure}
\epsfxsize=7cm
\centerline{\rotate[r]{\epsfbox{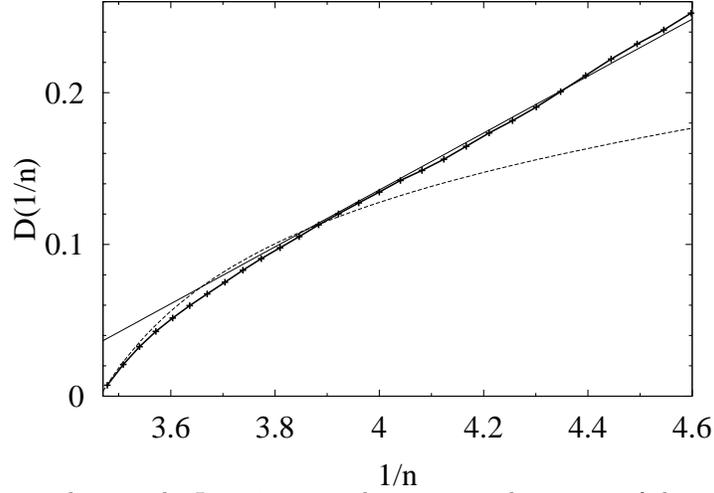}}}
\caption{Diffusion coefficient in the periodic Lorentz gas with
respect to the inverse of the number density of the scatterers
$n$. The Boltzmann approximation Eq.\ (\ref{eq:dbp}) is shown after
subtracting an offset of $\Delta D=0.025$ (thin bold line). It is
compared to the numerically exact results (thick bold line with
crosses), and to the Machta-Zwanzig approximation Eq.\ (\ref{eq:mz}).}
\label{fig7}
\end{figure}

\begin{figure}
\epsfxsize=7cm
\centerline{\rotate[r]{\epsfbox{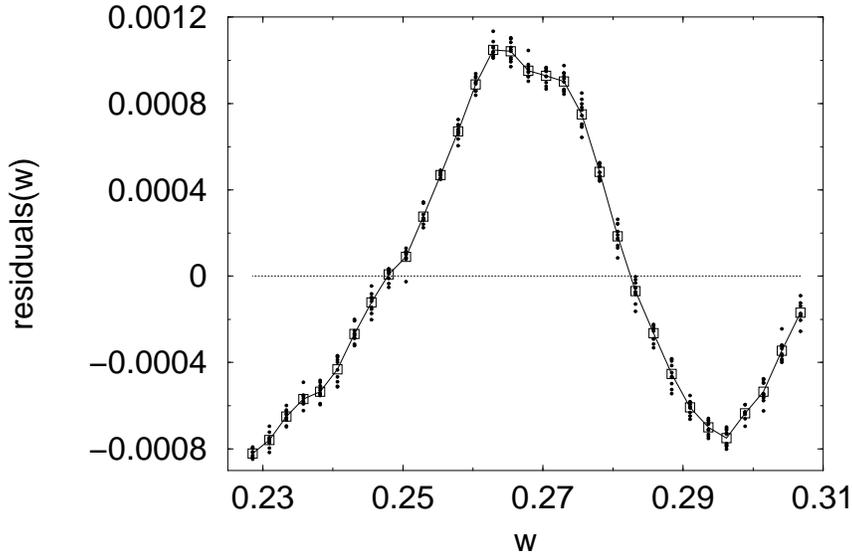}}}
\vspace*{1cm}
\caption{Residuals of $D)$, that is, the difference between the
numerically computed diffusion coefficient $D$ and a linear fit
over the range of the data presented in the figure. The dots
indicate the results of ten independent runs for each value of $w$, 
the squares represent the
algebraic averages over these ten runs. The size of the squares
is equal to the magnitude of the numerical error.}
\label{fig8}
\end{figure}

\begin{figure}
\epsfxsize=3cm
\epsfysize=7cm
\centerline{\rotate[r]{\epsfbox{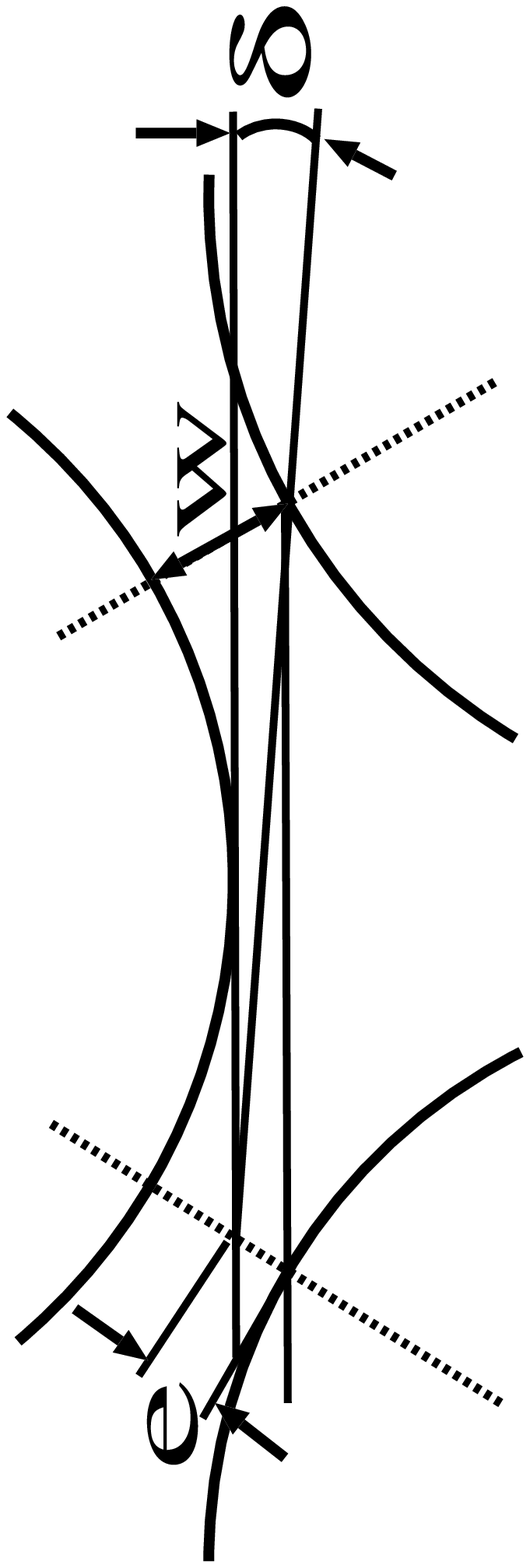}}}
  
\vspace*{1cm} 
\caption{Geometry to compute the probability of collisionless flights
$p_{\rm cf}$ across a trap. $e$ labels the size of one of the two
regions at any of the three entrances of the trap where ingoing
particles can perform collisionless flights. $\delta$ is related to an
approximate bound for the directions the velocity of such a particle
can take such that it performs a collisionless flight.}
\label{fig9}
\end{figure}

The corresponding second-order diffusion coefficient then reads
\begin{equation}
D_{3}=(16\tau)^{-1}[p_2l_2^2+p_{41}l_{41}^2+p_{42}l_{42}^2]
\quad . \label{eq:d3}
\end{equation}
Eq.\ (\ref{eq:d2}) thus corresponds to the second-order approximation
plotted in Fig.\ \ref{fig6} (a), and Eq.\ (\ref{eq:d3}) yields the
respective third-order approximation.

\end{appendix}

\newpage

\newpage

\newpage


\begin{thebibliography}{10}

\bibitem{GF2}
H. Fujisaka and S. Grossmann, Z. Physik B {\bf 48},  261  (1982).

\bibitem{GeNi82}
T. Geisel and J. Nierwetberg, Phys. Rev. Lett. {\bf 48},  7  (1982).

\bibitem{SFK}
M. Schell, S. Fraser, and R. Kapral, Phys. Rev. A {\bf 26},  504  (1982).

\bibitem{BaKl97}
E. Barkai and J. Klafter, Phys. Rev. Lett. {\bf 79},  2245  (1997).

\bibitem{Art1}
R. Artuso, Phys. Lett. A {\bf 160},  528  (1991).

\bibitem{Tse94}
H.-C. Tseng {\it et~al.}, Phys. Lett. A {\bf 195},  74  (1994).

\bibitem{Chen95}
C.-C. Chen, Phys. Rev. E {\bf 51},  2815  (1995).

\bibitem{Cvit95}
P. Cvitanovi\'{c}, lecture notes, available on www.nbi.dk/ChaosBook/ .

\bibitem{RKD}
R. Klages and J.R. Dorfman, Phys. Rev. Lett. {\bf 74},  387  (1995).

\bibitem{RKdiss}
R. Klages, {\em Deterministic diffusion in one-dimensional chaotic dynamical

  systems} (Wissenschaft \& Technik-Verlag, Berlin, 1996).

\bibitem{dcrc}
R. Klages and J.R. Dorfman, Phys. Rev. E {\bf 55},  R1247  (1997).

\bibitem{KlDo99}
R. Klages and J.R. Dorfman, Phys. Rev. E {\bf 59},  5361  (1999).

\bibitem{PG1}
P. Gaspard, J. Stat. Phys. {\bf 68},  673  (1992).

\bibitem{TG2}
S. Tasaki and P. Gaspard, J. Stat. Phys. {\bf 81},  935  (1995).

\bibitem{VTB97}
J. Vollmer, T. T\'el, and W. Breymann, Phys. Rev. Lett. {\bf 79},  2759
  (1997).

\bibitem{GaKl}
P. Gaspard and R. Klages, Chaos {\bf 8},  409  (1998).

\bibitem{DMP89}
I. Dana, N. Murray, and I. Percival, Phys. Rev. Lett. {\bf 65},  1693
(1989).

\bibitem{ReWi80}
A. Rechester and R. White, Phys. Rev. Lett. {\bf 44},  1586  (1980).

\bibitem{Lebo98}
P. Leboeuf, Physica D {\bf 116},  8  (1998).

\bibitem{Gasp}
P. Gaspard, {\em Chaos, Scattering, and Statistical Mechanics} (Cambridge
  University Press, Cambridge, 1998).

\bibitem{MaZw83}
J. Machta and R. Zwanzig, Phys. Rev. Lett. {\bf 50},  1959  (1983).

\bibitem{BarEC}
A. Baranyai, D.J. Evans, and E.G.D. Cohen, J. Stat. Phys. {\bf 70},
1085  (1993). 

\bibitem{MaMa97}
H. Matsuoka and R. Martin, J. Stat. Phys. {\bf 88},  81  (1997), 0022-4715.

\bibitem{CvGS92}
P. Cvitanovic, P. Gaspard, and T. Schreiber, Chaos {\bf 2},  85  (1992).

\bibitem{MoRo94}
G. Morriss and L. Rondoni, J. Stat. Phys. {\bf 75},  553  (1994).

\bibitem{GB1}
P. Gaspard and F. Baras, pp.\ 301--322 of Ref.\ \cite{MaHo92}.

\bibitem{GaBa95}
P. Gaspard and F. Baras, Phys.Rev.E {\bf 51},  5332  (1995).

\bibitem{MaHo92}
 {\em Microscopic simulations of complex hydrodynamic phenomena},
 Vol.~292 of {\em NATO ASI Series B: Physics}, edited by M. Mareschal
 and B.~L.  Holian (Plenum Press, New York, 1992).

\bibitem{TGN98}
 {\em Chaos and Irreversibility}, Vol.~8 of {\em Chaos}, edited by
 T.T\'el, P.Gaspard, and G.Nicolis (American Institute of Physics,
 College Park, 1998).

\bibitem{DoVL}
J.R. Dorfman, {\em An Introduction to Chaos in Nonequilibrium Statistical
  Mechanics} (Cambridge University Press, Cambridge, 1999).

\bibitem{BuSp}
L. Bunimovich and H. Spohn, Commun. Math. Phys. {\bf 176},  661  (1996).

\bibitem{BuSi1}
L. Bunimovich and Y. Sinai, Commun. Math. Phys. {\bf 78},  247  (1980).

\bibitem{BuSi2}
L. Bunimovich and Y. Sinai, Commun. Math. Phys. {\bf 78},  479  (1980).

\bibitem{WvL}
A. Weijland and J. van Leeuwen, Physica {\bf 38},  35  (1968).

\bibitem{Bruin}
C. Bruin, Physica {\bf 72},  261  (1974).

\bibitem{ChCo}
S. Chapman and T. Cowling, {\em The mathematical theory of non-uniform
gases}
  (Cambridge University Press, Cambridge, 1970).

\bibitem{CoRo98}
E. Cohen and L. Rondoni, Chaos {\bf 8},  357  (1998).

\bibitem{Ch2}
N. Chernov, C. Eyink, J. Lebowitz, and Y. Sinai, Comm. Math. Phys. {\bf
154},
  569  (1993).

\end{thebibliography}
\end{document}